\documentstyle[preprint,aps,epsfig]{revtex}
\begin{document}
\draft
\title{Nonlinear ER effects in an ac applied field}
\author{Jones T. K. Wan(1), G. Q. Gu(1,2) and K. W. Yu(1)}
\address{(1)Department of Physics, The Chinese University of Hong Kong,\\
         Shatin, New Territories, Hong Kong, China\\
	 (2)School of Computer Engineering,
	 University of Shanghai for Science and Technology,\\
	 Shanghai, 200 093 China}
\maketitle

\begin{abstract}
The electric field used in most electrorheological (ER)
experiments is usually quite high, and nonlinear ER effects have
been theoretically predicted and experimentally measured recently.
A direct method of measuring the nonlinear ER effects is to
examine the frequency dependence of the same effects. For a
sinusoidal applied field, we calculate the ac response which
generally includes higher harmonics.
In is work, we develop a multiple image formula, and calculate the
total dipole moments of a pair of dielectric spheres, embedded in
a nonlinear host. The higher harmonics due to the nonlinearity are
calculated systematically.
\end{abstract}
\vskip 5mm

\section{Introduction}
Electrorheological (ER) fluids consist of highly polarizable
particles in a nearly insulating fluid. Upon the application of
electric fields, the apparent viscosity of ER fluids can be
changed by several orders of magnitude, due to the formation of
chains of and columns of particles across the electrodes in the
direction of the applied field. The rapid transition between the
fluid and solid phases renders this material the potential of
important technological applications.

On the other hand, the applied electric field used in most ER experiments is
usually quite high, and important data on nonlinear ER effects induced by a
strong electric field have been reported by Klingenberg and coworkers
\cite{Kling98}. Recently, the effect of a nonlinear characteristic on the
interparticle force has been analyzed in an ER suspension of nonlinear
particles \cite{nler-1} and further extended to a nonlinear host medium
\cite{nler-2}. These work confirmed the previous theoretical results that the
attractive force between two touching spheres can have a quasi-linear
dependence \cite{Felici}.

A convenient method of probing the nonlinear characteristic is to measure the
harmonics of the induced polarization under the application of a sinusoidal
(ac) electric field \cite{Kling98}. When a nonlinear composite with nonlinear
dielectric particles embedded in a host medium, or with a nonlinear host medium
is subjected to a sinusoidal field, the electrical response in the composite
will in general be a superposition of many sinusoidal functions
\cite{Bergman95}. It is natural to investigate the effects of a nonlinear
characteristic on the interparticle force in an ER fluid which can be regarded
as a nonlinear composite medium \cite{Gu00}. The strength of the nonlinear
polarization is reflected in the magnitude of the harmonics.

In this work, we will develop a self-consistent theory to
calculate the ac response of a nonlinear ER fluid. Our theory goes
beyond the simple point-dipole approximation and accounts for the
multipole interaction between the polarized particles.

\section{Nonlinear polarization and its higher harmonics}
We first consider an isolated spherical particle of radius $a$ and
dielectric constant $\epsilon_p$. The sphere is placed in a host
medium of dielectric constant $\epsilon_m$. When a time-dependent
electric field ${\bf E}=E(t)\hat{z}$ is applied. The particle will
be polarized and its surface charge will contribute to it a dipole
moment $p_{0}=\epsilon_m a^3 \beta E(t)$, where $\beta$ is the
dipolar factor given by:
$\beta=(\epsilon_p-\epsilon_m)/(\epsilon_p+2\epsilon_m)$. In ER
fluids, when the particles get close to one another, they do not
behave as point dipole and we have to consider the mutual
polarization between the particles.

Next, we examine the effect of a nonlinear characteristic on the
induced dipole moment. We concentrate on the case where the host
medium have a nonlinear dielectric constant, while the suspended
particles are linear. The nonlinear characteristic gives rise to a
field-dependent dielectric coefficient. In other words, the
electric displacement-electric field relation in the host medium
is given by:
\begin{equation}
{\bf D}_m=\epsilon_m{\bf E}_m +\chi_m\langle E_m^2\rangle{\bf E}_m
  =\widetilde{\epsilon}_m{\bf E}_m,
\label{nl-coefficient-host}
\end{equation}
where $\chi_m$ is the nonlinear coefficient of the particles. We
have assumed that the average field inside the host medium is
uniform. This is called the decoupling approximation \cite{Yu96}.
It has been shown that such an approximation yields a lower bound
for the accurate result \cite{Yu96}. Furthermore, in ER fluids the
frequency of the applied field is low; we can assume both
$\epsilon$ and $\chi$ are independent of frequency. As a result,
the induced dipole moment is field-dependent and is given by:
\begin{equation}
\widetilde{p}_{0}=\widetilde{\epsilon}_m a^3\widetilde{\beta}E(t),
\end{equation}
where $\widetilde{\beta}$ is the field-dependent dipolar factor
and is given by:
\begin{equation}
\widetilde{\beta}={\epsilon_p-\widetilde{\epsilon}_m \over
\epsilon_p+2\widetilde{\epsilon}_m}=
{\epsilon_p-(\epsilon_m+\chi_m\langle E_m^2\rangle) \over
\epsilon_p+2(\epsilon_m+\chi_m\langle E_m^2\rangle)}.
\end{equation}

When the polarized spheres approach to one another, they will be
further polarized by the mutual polarization effect. As a result,
the point-dipole approximation breaks down and we must consider
the multipole moments.

Let us consider a pair of nonlinear dielectric spheres of the same
radius $a$, separated by a distance $r$. Each of them has a
dielectric coefficient $\epsilon_p$. By using the method of
multiple images \cite{Yu,Poladian}, we deduce the total dipole
moment of the spheres:
\begin{equation}
\widetilde{p}_{\rm T} =
\widetilde{p}_0\sum_{n=0}^\infty(-\widetilde{\beta})^n
\left({\sinh\alpha\over\sinh (n+1)\alpha} \right)^3.
\label{trans-dielectric}
\end{equation}
The subscript $T$ denotes that the applied electric field is
perpendicular to the line joining the centers of the particles
(i.e., a transverse field). The parameter $\alpha$ is related to
the separation between the particles: $\cosh\alpha=r/2a=\sigma$,
where $\sigma$ is the reduced separation. For a longitudinal
field, we change the factor from $(-\widetilde{\beta})$ to
$(2\widetilde{\beta})$. The above formula can be generalized to
account for a pair of dielectric spheres of the different
sizes \cite{Yu}. The multiple images method was widely adopted
\cite{nler-1,nler-2,dyer} to calculate the total dipole moment as
well as the interparticle force.

We should remark that the present multiple images method is an
approximation only. In fact there is a more complicated images
method for a dielectric sphere \cite{Choy1,Choy2}, which gives the
exact image dipole moment of a dielectric sphere that placed in
front of a point dipole. However, the numerical results of the
present version of formula agree with the numerical solution of an
integral equation method \cite{etopim1,etopim2}.

When we apply a sinusoidal electric field, i.e., $E(t)=E_0 \sin
{\omega t}$, the induced dipole moment will vary with time
sinusoidally. Due to the nonlinearity of the particles, the
induced dipole moment will be a superposition of harmonics.  In
other words, we have:
\begin{equation}
\widetilde{p}_{\rm T} = p_{\omega} \sin {\omega t}
 +p_{3\omega} \sin {3\omega t}
 +p_{5\omega} \sin {5\omega t} + \cdots
\label{dipole-expand}
\end{equation}
Due to the inversion symmetry of the dielectric media, only the
harmonics of odd order survive. The coefficient $p_{n\omega}$ of
the nth harmonic is given by:
\begin{equation}
p_{n\omega} = {\omega\over \pi}\int_0^{2\pi/\omega} \widetilde{p}_{\rm T}\sin{n\omega
t} dt.
\label{dipole-harmonics}
\end{equation}

In what follows, we report results for the transverse field case
only. The longitudinal field case is similar. In the next section,
we will use the series expansion to obtain analytic expressions
for the harmonics of the induced dipole moment. We will obtain the
coefficient $p_{n\omega}$ as a power series of the applied field
$E_0$.

\section{Self-consistent evaluation of the average field}
According to Eqs.(\ref{trans-dielectric}), the induced dipole
moment of the dielectric spheres $\widetilde{p}_{\rm T}$ can be
determined if we find the average field $\langle E_m^2\rangle$.
The electric field inside the host medium can be conveniently
calculated by considering the effective nonlinear dielectric
constant of a two-component composite. For a two-component
composite, the effective nonlinear dielectric constant
$\widetilde{\epsilon}_e$ is given by \cite{Yu93}:
\begin{eqnarray}
\widetilde{\epsilon}_e = {1\over E^2(t) V}\int_V
 \epsilon({\bf r})|{\bf E}({\bf r},t)|^2dV
  = {f \epsilon_p\over E^2(t)}\langle E_p^2\rangle
  + {(1-f) \widetilde{\epsilon}_m\over E^2(t)}\langle E_m^2\rangle,
\end{eqnarray}
where $f$ is the volume fraction of the particles. For a pair of
spheres inside a transverse field, the effective nonlinear
Dielectric constant can be expressed as \cite{Yu}:
\begin{equation}
\tilde{\epsilon_e}=\widetilde{\epsilon}_m +3 f
\widetilde{\epsilon}_m\left ({ \widetilde{\beta}
\widetilde{p}_{\rm T}\over \widetilde{p}_0}\right ).
\label{effective-e}
\end{equation}
The average field of the host medium can be calculated by using
Eq.(\ref{effective-e}):
\begin{eqnarray}
\langle E_m^2\rangle = {1\over 1-f}E^2(t)
  {\partial \widetilde{\epsilon}_e\over\partial\widetilde{\epsilon}_m}.
\label{solve-local-field}
\end{eqnarray}
For a nonlinear characteristic [Eq.(\ref{nl-coefficient-host})],
we solve Eq.(\ref{solve-local-field}) self-consistently
\cite{Yu96}. The average field inside the host medium as well as
the dipole moment of the spheres [Eq.(\ref{trans-dielectric})] can
be determined.

It remains to examine how the higher harmonics of
$\widetilde{p}_{\rm T}$ depends the nonlinearity. We first
normalize the material parameters with the linear dielectric
coefficient of the host $\epsilon_m$:
\begin{equation}
\epsilon_p'={\epsilon_p\over\epsilon_m},\quad
\chi_m'={\chi_m\over\epsilon_m},\quad
\widetilde{\epsilon}_m'={\widetilde{\epsilon}_m\over\epsilon_m}=1+\chi_m'\langle
E_m^2\rangle.
\end{equation}
We then expand $\langle E_m^2\rangle$ into a Taylor expansion:
\begin{equation}
\chi_m'\langle E_m^2\rangle ={\chi_m' E^2(t)\over 1-f} + {3f\over
1-f} \chi_m' E^2(t)\sum_{s=0}^\infty
c_s\left(\chi_m'\langle E_m^2\rangle\right)^s,
\label{local-field-taylor}
\end{equation}
where the expansion coefficient $c_s$ is given by:
\begin{equation}
c_s={1\over s!}
{\partial^{s+1}\over\partial\widetilde{\epsilon'}_m^{s+1}}
\left[\widetilde{\epsilon}'_m\widetilde{\beta}\sum_{n=0}^\infty(-\widetilde{\beta})^n
\left({\sinh\alpha\over\sinh (n+1)\alpha}
\right)^3\right]_{\widetilde{\epsilon}'_m=1}.
\end{equation}
Hence the expansion coefficients do not depend on the applied field. Similarly,
we expand $\widetilde{p}_{\rm T}$ into a Taylor expansion:
\begin{equation}
\widetilde{p}_{\rm T} = \epsilon_m a^3E(t)\sum_{s=0}^\infty
a_s\left(\chi_m'\langle E_m^2\rangle\right)^s,
\label{dipole-taylor}
\end{equation}
where
\begin{equation}
a_s={1\over s!}{\partial^s\over\partial\widetilde{\epsilon'}_m^s}
\left[\widetilde{\epsilon}_m'
\widetilde{\beta}\sum_{n=0}^\infty(-\widetilde{\beta})^n
\left({\sinh\alpha\over\sinh (n+1)\alpha}
\right)^3\right]_{\widetilde{\epsilon}_m'=1}.
\end{equation}
It is easy to show that: $c_s=(s+1)a_{s+1}$. In the case of a weak
nonlinearity, i.e. $\chi_m' E^2(t)\ll1$, we can rewrite
Eqs.(\ref{local-field-taylor}) and (\ref{dipole-taylor}), keeping
only the lowest orders of $\chi_m' E^2(t)$ and $\chi_m'\langle
E_m^2\rangle$:
$$
\chi_m'\langle E_m^2\rangle = {\chi_m' E^2(t)\over 1-f}+{3f\over
1-f} \chi_m' E^2(t)
 (c_0 + c_1\chi_m'\langle E_m^2\rangle + \cdots).
$$
The first order term of $\chi_m' E^2(t)$ gives the average field
inside a linear dielectric host medium. Similarly, the induced
dipole moment is given by:
\begin{eqnarray}
\widetilde{p}_{\rm T} &=& \epsilon_m a^3a_0 E(t) + \epsilon_m
a^3a_1{\chi_m' E^3(t)\over 1-f} + {3f\over 1-f}\epsilon_m a^3
a_1^2\chi_m' E^3(t) +\cdots \nonumber\\
&=& K_1 E(t)+K_3 E^3(t) + \cdots.
\label{dipole-lower-order}
\end{eqnarray}
It should be remarked
that, from Eq.(\ref{dipole-taylor}), $K_1E(t)$ is the linear
dipole moment $p_{\rm T}$:
\begin{equation}
p_{\rm T}=\epsilon_m a^3 \beta E(t)\sum_{n=0}^\infty(-\beta)^n
\left({\sinh\alpha\over\sinh (n+1)\alpha} \right)^3.
\end{equation}
Let us consider a sinusoidal applied electric field $E(t)=E_0 \sin {\omega t}$.
By using the identity $4\sin^3{\omega t} = 3\sin{\omega t}-\sin{3\omega t}$, we
can expand $E^3(t)$ in terms of the first and the third harmonics. By comparing
Eq.(\ref{dipole-expand}) with Eq.(\ref{dipole-lower-order}), we find:
$$
p_\omega = K_1 E_0+{3\over4}K_3 E_0^3 \quad{\rm and}\quad p_{3\omega} =
-{1\over 4}K_3 E_0^3.
$$
The above results show that the induced dipole moment must include the higher
harmonics \cite{Gu00}. Furthermore, the results show that $p_\omega$ also
depends on $E_0^3$. This is a nontrivial result as it implies that the first
harmonic of the induced dipole moment depends on the strength of the
nonlinearity. Concomitantly, the higher harmonics should become more
significant as $E_0$ gets higher. In the case of a higher applied field, we
must include even higher order terms [i.e., higher powers of $E(t)$] in the
expansion of $\widetilde{p}_{\rm T}$:
$$
\widetilde{p}_{\rm T} = K_1 E(t)+K_3 E^3(t)+K_5 E^5(t) +\cdots.
$$
Again, by considering the identity $16\sin^5{\omega t}=10\sin{\omega t}
-5\sin{3\omega t}+\sin{5\omega t}$, the harmonics are given by:
\begin{eqnarray*}
p_\omega &=& K_1E_0+{3\over4}K_3E_0^3+{10\over 16}K_5 E_0^5,\\
p_{3\omega} &=& -{1\over 4}K_3E_0^3-{5\over 16}K_5 E_0^5,\\
p_{5\omega} &=& {1\over 16}K_5E_0^5.
\end{eqnarray*}

Consequently, in the case of nonlinear ac response, the
convergence of the series expansion is questionable and a
self-consistent formalism is needed. The self-consistent equation
[Eq.(\ref{solve-local-field})] can help to solve for the average
field. From Eq.(\ref{local-field-taylor}), (\ref{dipole-taylor})
and (\ref{dipole-lower-order}), we find that:
\begin{equation}
{\widetilde{p}_{\rm T}\over p_0}=F(\chi_m' E_0^2) \quad{\rm
and}\quad \chi_m'\langle E_m^2\rangle = G(\chi_m' E_0^2),
\label{power-law}
\end{equation}
where $F$ and $G$ are functions of a single variable, and
$p_0=\epsilon_m a^3\beta E_0$; this is the linear dipole moment of
an isolated dielectric particle under a time-independent applied
field $E_0$. Similar conclusion can be drawn for $p_\omega$ and
$p_{3\omega}$. This demonstrates that we can use $\chi_m'E_0^2$ as
a natural variable in the nonlinear composite problem \cite{Gu92}.

\section{Numerical Results}
In this section, we perform numerical calculations to investigate
the effects of a nonlinear characteristic on the harmonics of the
induced dipole moment and the average electric field. As shown in
section III, the induced dipole moment of the dielectric spheres
is affected by three factors: the strength of the applied field
$(E_0)$, the relative nonlinear dielectric coefficient of the host
medium $(\chi_m')$ and the relative linear dielectric constants of
the particles $(\epsilon_p')$. In order to emphasize the effect of
mutual polarization, we use a small reduced separation
$\sigma=1.1$. From Eq.(\ref{power-law}), we can use $\chi_m'
E_0^2$ as the variable to plot the numerical results.

Before showing the numerical results, we consider the simple case
of an isolated sphere. In this case, the average field inside the
host medium can be solved exactly \cite{nler-2,Gu92}:
\begin{equation}
\chi_m'\langle E_m^2\rangle = \left({1\over 1-f}+ {3f\over 1-f}
 {\epsilon_p^2-2\epsilon_p\widetilde{\epsilon}_m-2\widetilde{\epsilon}_m^2
 \over(\epsilon_p+2\widetilde{\epsilon}_m)^2}\right)\chi_m'E^2(t).
\end{equation}
When the applied field is small, the effect of nonlinearity is
negligible. The average field is dominated by the first part of
the above equation. Since the volume fraction $f$ have to be
small \cite{Yu93}, the average field can be well approximated by:
\begin{equation}\label{average-field-approx}
\langle E_m^2\rangle\approx E^2(t).
\end{equation}
For a sinusoidal applied field $E(t)=E_0\sin\omega t$, we have:
$$
E_\omega \approx E_0,
$$
which means that the first harmonic varies linearly with $E_0$. In other words,
a linear relation between $E_\omega$ and $E_0$ is an indication of a weak
nonlinearity.

In Fig.\ref{acer-host-field}, we plot the first and the third
harmonics of the average field, with $\epsilon_p'=2$ $(\beta=1/4)$
and $\epsilon_p'=10$ $(\beta=1/4)$, respectively. The volume
fraction has a small influence on the nonlinear effects. The
harmonics are plotted against $\sqrt{\chi_m'} E_0$. The first
harmonic $\sqrt{\chi_m'}E_\omega$ varies almost linearly with
$\sqrt{\chi_m'}E_0$. Moreover, the magnitude of the third
harmonics is small compared with the first harmonics. These
results show that the average field can be approximated by
Eq.(\ref{average-field-approx}).

The numerical results of the average field suggest that the
average field varies almost linearly with the applied field.
However, as suggested by Eq.(\ref{dipole-lower-order}),the dipole
moment does not vary linearly with the applied field. In
Fig.\ref{acer-host-dipole}, we plot the first and the third
harmonics of the induced dipole moment. We divide the harmonics by
$p_0=\epsilon_m a^3\beta E_0$, which is the linear dipole moment
of an isolated dielectric particle under a time-independent
applied field $E_0$. In Fig.\ref{acer-host-dipole}, we find that
the first harmonics changes from positive to negative. This is
because as the magnitude of the applied field increase, the
field-dependent dipolar factor $\widetilde{\beta}$ changes from
positive to negative. Hence the first harmonics changes its sign.

Next we examine the third harmonics of the dipole moment. The
ratio $p_{3\omega}/p_0$ increase with the applied field. The third
harmonics is an evidence of nonlinear effects. The ratio increase
significantly when the dielectric contrast between the particles
and the host is small. In other words, a smaller dielectric
contrast gives a larger nonlinear response.

\section*{Discussion and conclusion}

Here a few comments on our results are in order. In the present
study, we have examined the case of linear particles suspending in
a nonlinear host. The case of nonlinear particles suspending in a
linear host has already been studied by similar formalism
\cite{acer-pre}.

So far, we have not considered the frequency dependence of the
dielectric constants of the host medium and the particles. In a
realistic situation, the dielectric constants can decrease with
the increase of the frequency. For simplicity, we may adopt the
Debye relaxation expression.

As we have obtained the expression for the induced dipole moments, we may take
a step forward to calculate the interparticle force via the energy approach
\cite{Yu}. We will find a time-dependent force $F(t)$ but only its time average
could be measured during an experiment. From the energy approach \cite{Yu},
$$
\langle F_T(t) \rangle
 = {\partial \langle E(t) p_T (t)\rangle \over \partial r}.
$$
We believe that the interparticle force in the ac case should
differ significantly from that of the dc case of nonlinear ER
fluids because, as we have shown, the nonlinearity enters into the
composite problem in a nontrivial way. Results of the
interparticle force will be published elsewhere \cite{msm}.

In conclusion, we have considered the effects of a nonlinear
characteristic on the ER fluid under the influence of a sinusoidal
applied field. We have calculated the harmonic components of the
induced dipole moment as well as the average electric field. We
have also examined the conditions for obtaining large ac response
in ER fluids.

\section*{Acknowledgments}
This work was supported by the Research Grants Council of the Hong Kong SAR
Government under project number CUHK 4284/00P. G. Q. Gu acknowledges financial
support from the Key Project of the National Natural Science Foundation of
China under project number 19834070.

\begin{figure}[h]
\caption{The first and the third harmonics of the average field,
with $\epsilon_p'=2$ $(\beta=1/4)$ and $\epsilon_p'=10$
$(\beta=1/4)$, respectively. The volume fraction has a small
influence on the nonlinear effects.}
\label{acer-host-field}
\end{figure}

\begin{figure}[h]
\caption{The first and the third harmonics of the induced dipole
moment.  The material parameters are chosen as the same as in
Fig.\ref{acer-host-field}. The nonlinear effects are significant
when the dielectric contrast is small.}
\label{acer-host-dipole}
\end{figure}

\newpage
\centerline{\epsfig{file=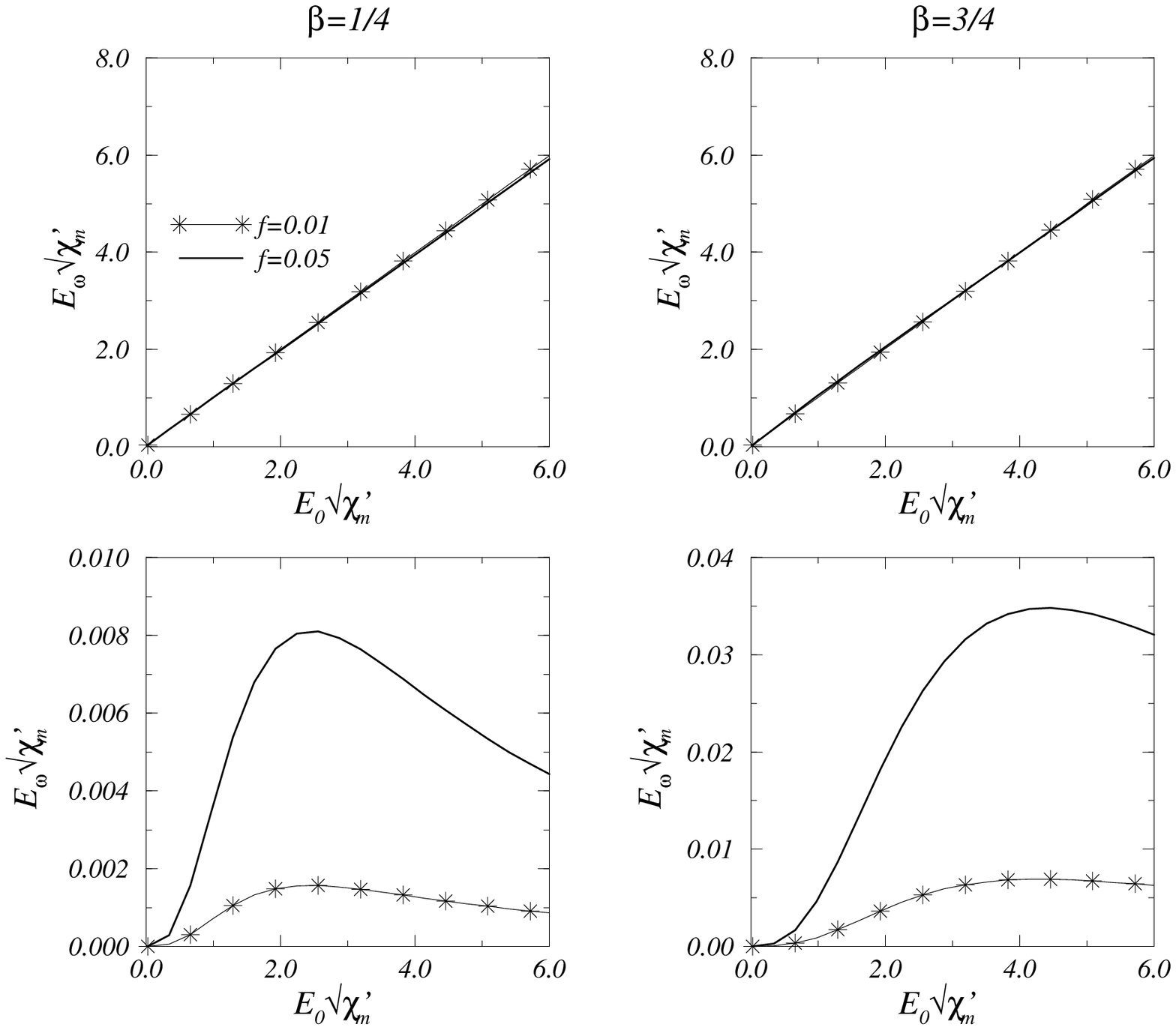,width=\linewidth}}

\newpage
\centerline{\epsfig{file=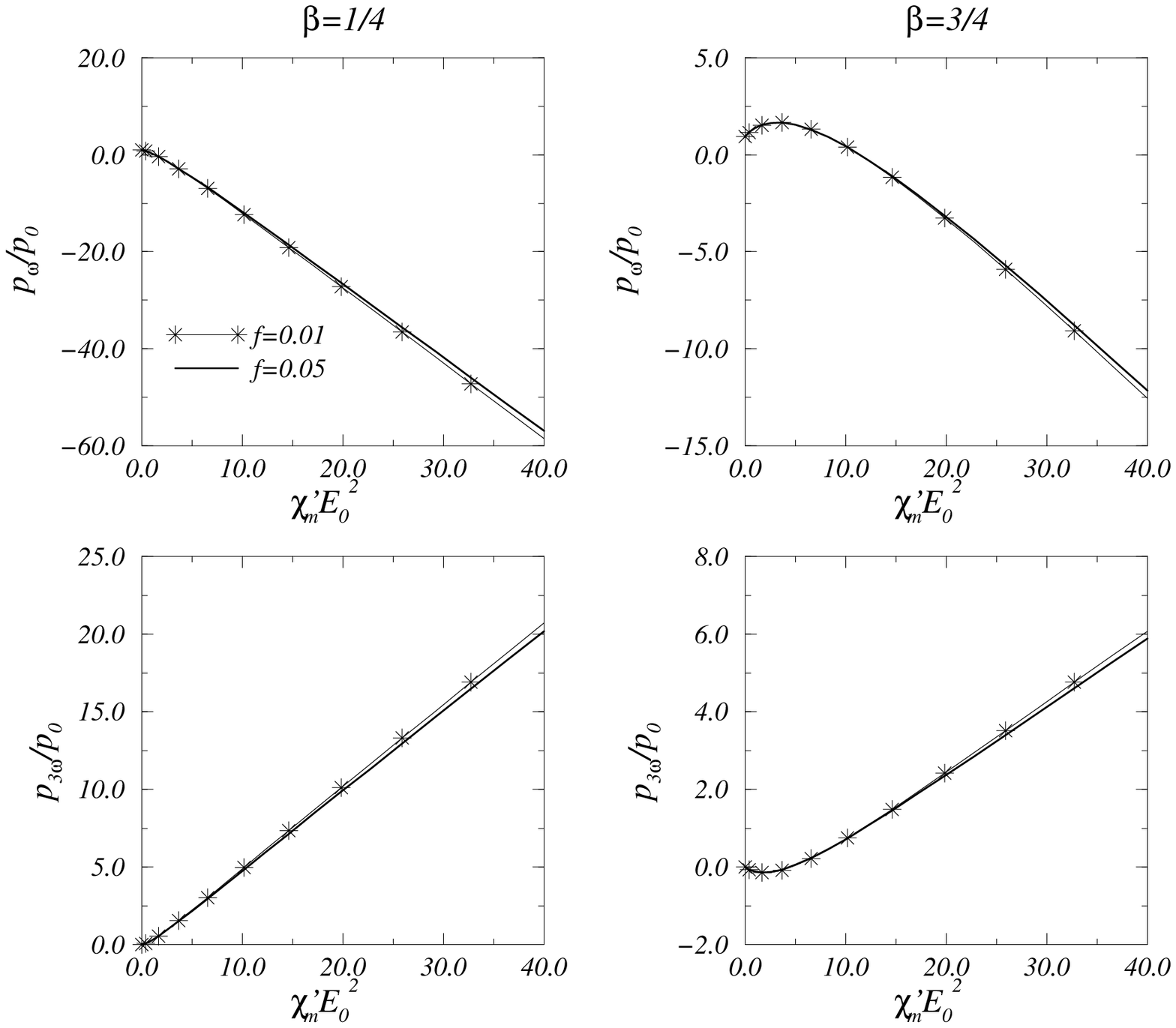,width=\linewidth}}

\end{document}